\documentclass[preprint2]{aastex}
\usepackage{graphicx}
\usepackage{epstopdf}
\usepackage{dcolumn}
\usepackage{bm}
\usepackage{natbib}
\usepackage{hyperref}

\def\nue{$\nu_e$}
\def\nuebar{$\bar{\nu}_e$}

\def\heavywater{D$_2$O}
\def\water {H$_2$O}
\newcommand{\iso}[2]{{}^{#1}{\rm #2}}

\begin{document}
\newcommand{\alta}{Department of Physics, University of 
Alberta, Edmonton, Alberta, T6G 2R3, Canada}
\newcommand{\ubc}{Department of Physics and Astronomy, University of 
British Columbia, Vancouver, BC V6T 1Z1, Canada}
\newcommand{\bnl}{Chemistry Department, Brookhaven National 
Laboratory,  Upton, NY 11973-5000}
\newcommand{\carleton}{Ottawa-Carleton Institute for Physics, Department of Physics, Carleton University, Ottawa, Ontario K1S 5B6, Canada}
\newcommand{\uog}{Physics Department, University of Guelph,  
Guelph, Ontario N1G 2W1, Canada}
\newcommand{\lu}{Department of Physics and Astronomy, Laurentian 
University, Sudbury, Ontario P3E 2C6, Canada}
\newcommand{\lbnl}{Institute for Nuclear and Particle Astrophysics and 
Nuclear Science Division, Lawrence Berkeley National Laboratory, Berkeley, CA 94720}
\newcommand{\lbla}{ Lawrence Berkeley National Laboratory, Berkeley, CA}
\newcommand{\lanl}{Los Alamos National Laboratory, Los Alamos, NM 87545}
\newcommand{\llnl}{Lawrence Livermore National Laboratory, Livermore, CA}
\newcommand{\lanla}{Los Alamos National Laboratory, Los Alamos, NM 87545}
\newcommand{\oxford}{Department of Physics, University of Oxford, 
Denys Wilkinson Building, Keble Road, Oxford OX1 3RH, UK}
\newcommand{\penn}{Department of Physics and Astronomy, University of 
Pennsylvania, Philadelphia, PA 19104-6396}
\newcommand{\queens}{Department of Physics, Queen's University, 
Kingston, Ontario K7L 3N6, Canada}
\newcommand{\uw}{Center for Experimental Nuclear Physics and Astrophysics, 
and Department of Physics, University of Washington, Seattle, WA 98195}
\newcommand{\uta}{Department of Physics, University of Texas at Austin, Austin, TX 78712-0264}
\newcommand{\triumf}{TRIUMF, 4004 Wesbrook Mall, Vancouver, BC V6T 2A3, Canada}
\newcommand{\ralimp}{Rutherford Appleton Laboratory, Chilton, Didcot OX11 0QX, UK}
\newcommand{\iusb}{Department of Physics and Astronomy, Indiana University, South Bend, IN}
\newcommand{\fnal}{Fermilab, Batavia, IL}
\newcommand{\uo}{Department of Physics and Astronomy, University of Oregon, Eugene, OR}
\newcommand{\hu}{Department of Physics, Hiroshima University, Hiroshima, Japan}
\newcommand{\slac}{Stanford Linear Accelerator Center, Menlo Park, CA}
\newcommand{\mac}{Department of Physics, McMaster University, Hamilton, ON}
\newcommand{\doe}{US Department of Energy, Germantown, MD}
\newcommand{\lund}{Department of Physics, Lund University, Lund, Sweden}
\newcommand{\mpi}{Max-Planck-Institut for Nuclear Physics, Heidelberg, Germany}
\newcommand{\uom}{Ren\'{e} J.A. L\'{e}vesque Laboratory, Universit\'{e} de Montr\'{e}al, Montreal, PQ}
\newcommand{\cwru}{Department of Physics, Case Western Reserve University, Cleveland, OH}
\newcommand{\pnnl}{Pacific Northwest National Laboratory, Richland, WA}
\newcommand{\uc}{Department of Physics, University of Chicago, Chicago, IL}
\newcommand{\mitt}{Laboratory for Nuclear Science, Massachusetts Institute of Technology, Cambridge, MA 02139}
\newcommand{\ucsd}{Department of Physics, University of California at San Diego, La Jolla, CA }
\newcommand{	\lsu	}{Department of Physics and Astronomy, Louisiana State University, Baton Rouge, LA 70803}
\newcommand{\imp}{Imperial College, London SW7 2AZ, UK}
\newcommand{\uci}{Department of Physics, University of California, Irvine, CA 92717}
\newcommand{\ucia}{Department of Physics, University of California, Irvine, CA}
\newcommand{\suss}{Department of Physics and Astronomy, University of Sussex, Brighton  BN1 9QH, UK}
\newcommand{\lifep}{Laborat\'{o}rio de Instrumenta\c{c}\~{a}o e F\'{\i}sica Experimental de
Part\'{\i}culas, Av. Elias Garcia 14, 1$^{\circ}$, 1000-149 Lisboa, Portugal}
\newcommand{\hku}{Department of Physics, The University of Hong Kong, Hong Kong.}
\newcommand{\aecl}{Atomic Energy of Canada, Limited, Chalk River Laboratories, Chalk River, ON K0J 1J0, Canada}
\newcommand{\nrc}{National Research Council of Canada, Ottawa, ON K1A 0R6, Canada}
\newcommand{\princeton}{Department of Physics, Princeton University, Princeton, NJ 08544}
\newcommand{\birkbeck}{Birkbeck College, University of London, Malet Road, London WC1E 7HX, UK}
\newcommand{\snoi}{SNOLAB, Sudbury, ON P3Y 1M3, Canada}
\newcommand{\uba}{University of Buenos Aires, Argentina}
\newcommand{\hvd}{Department of Physics, Harvard University, Cambridge, MA}
\newcommand{\pny}{Goldman Sachs, 85 Broad Street, New York, NY}
\newcommand{\pnv}{Remote Sensing Lab, PO Box 98521, Las Vegas, NV 89193}
\newcommand{\psis}{Paul Schiffer Institute, Villigen, Switzerland}
\newcommand{\liverpool}{Department of Physics, University of Liverpool, Liverpool, UK}
\newcommand{\uto}{Department of Physics, University of Toronto, Toronto, ON, Canada}
\newcommand{\uwisc}{Department of Physics, University of Wisconsin, Madison, WI}
\newcommand{\psu}{Department of Physics, Pennsylvania State University,
     University Park, PA}
\newcommand{\anl}{Deparment of Mathematics and Computer Science, Argonne
     National Laboratory, Lemont, IL}
\newcommand{\cornell}{Department of Physics, Cornell University, Ithaca, NY}
\newcommand{\tufts}{Department of Physics and Astronomy, Tufts University, Medford, MA}
\newcommand{\ucd}{Department of Physics, University of California, Davis, CA}
\newcommand{\unc}{Department of Physics, University of North Carolina, Chapel Hill, NC}
\newcommand{\dresden}{Institut f\"{u}r Kern- und Teilchenphysik, Technische Universit\"{a}t Dresden,  01069 Dresden, Germany}
\newcommand{\isu}{Department of Physics, Idaho State University, Pocatello, ID}
\newcommand{\qmul}{Dept. of Physics, Queen Mary University, London, UK}
\newcommand{\ucsb}{Dept. of Physics, University of California, Santa Barbara, CA}
\newcommand{\cern}{CERN, Geneva, Switzerland}
\newcommand{\utah}{Dept. of Physics, University of Utah, Salt Lake City, UT}
\newcommand{\casa}{Center for Astrophysics and Space Astronomy, University
of Colorado, Boulder, CO 80309}
\newcommand{\susel}{Sanford Underground Research Laboratory, Lead, SD}  
\newcommand{\ntu}{Center of Cosmology and Particle Astrophysics, National Taiwan University, Taiwan}
\newcommand{\berlin}{Institute for Space Sciences, Freie Universit\"{a}t Berlin,
Leibniz-Institute of Freshwater Ecology and Inland Fisheries, Germany}
\newcommand{\bhsu}{Black Hills State University, Spearfish, SD} 
\newcommand{\queensa}{Dept.\,of Physics, Queen's University, 
Kingston, Ontario, Canada} 
\newcommand{\aasu}{Dept.\,of Chemistry and Physics, Armstrong Atlantic State University, Savannah, GA}
\newcommand{\ucb}{Physics Department, University of California at Berkeley, and Lawrence Berkeley National Laboratory, Berkeley, CA}
\newcommand{\mcgill}{Physics Department, McGill University, Montreal, QC, Canada}
\newcommand{\columbia}{Columbia University, New York, NY}
\newcommand{\rhul}{Dept. of Physics, Royal Holloway University of London, Egham, Surrey, UK}
\newcommand{\ubama}{Department of Physics and Astronomy, University of Alabama, Tuscaloosa, AL}
\newcommand{\kit}{Instit\"{u}t f\"{u}r Experimentelle Kernphysik, Karlsruher Instit\"{u}t f\"{u}r Technologie, Karlsruhe, Germany}
\newcommand{\uwinn}{Department of Physics, University of Winnipeg, Winnipeg, MB  R3B 2E9, Canada}
\newcommand{\sju}{Department of Physics, Shanghai Jiaotong University, Shanghai, China}

\author{B.~Aharmim\altaffilmark{6},
S.\,N.~Ahmed\altaffilmark{14},
A.\,E.~Anthony\altaffilmark{17, a},
N.~Barros\altaffilmark{8},
E.\,W.~Beier\altaffilmark{13},
A.~Bellerive\altaffilmark{4},
B.~Beltran\altaffilmark{1},
M.~Bergevin\altaffilmark{7,5,b},
S.\,D.~Biller\altaffilmark{12},
K.~Boudjemline\altaffilmark{4, 14}
% ...
}
\author{SNO Collaboration}

\altaffiltext{1}{\alta}
% ...
\altaffiltext{4}{\carleton}
\altaffiltext{5}{\uog}
\altaffiltext{6}{\lu}
\altaffiltext{7}{\lbnl}
\altaffiltext{8}{\lifep}
% ...
\altaffiltext{12}{\oxford}
\altaffiltext{13}{\penn}
\altaffiltext{14}{\queens}
% ...
\altaffiltext{17}{\uta}
% ...
\altaffiltext{a}{Present address: \casa}
\altaffiltext{b}{Present address: \ucd}
\altaffiltext{c}{Present address: \kit}
% ...

\title{A Search for Astrophysical Burst Signals at the Sudbury Neutrino Observatory}


\begin{abstract}
The Sudbury Neutrino Observatory (SNO) has confirmed the standard solar model and neutrino oscillations through the observation of neutrinos from the solar core. In this paper we present a search for neutrinos associated with sources other than the solar core, such  as gamma-ray bursters and solar flares. We present a new method for looking for temporal coincidences between neutrino events and astrophysical bursts of widely varying intensity. No correlations were found between neutrinos detected in SNO and such astrophysical sources.
\end{abstract}

\maketitle

\section{Introduction} 
\label{sec:intro}

The Sudbury Neutrino Observatory (SNO) collaboration has looked for time-dependent anomalies due to both periodic variations~\citep{Aharmim:2005iu,Collaboration:2009qz} in neutrino flux and short bursts of neutrinos~\citep{Collaboration:2010gx}. In the present study, we are specifically searching for neutrino events that are correlated to other known astrophysical events.  There is a wide variety of potential astrophysical sources of neutrinos. In this paper we consider ``burst" events, which include $\gamma $-ray bursters (GRB's), solar flares, magnetars, and an intense burst observed in the Parkes radio telescope. These astrophysical burst events are short-lived and occur at random. It is expected that any related neutrino signal would be undetectable~\citep{Piran:2004ba, Bahcall:1988vz} and the present experimental constraints provide only limits~\citep{Hirata:1988sx, Hirata_90, Aglietta:1991fv, Fukuda:2002nf, Thrane:2009fj}.

Similar searches for neutrinos temporally correlated with astrophysical events have  treated all bursts (regardless of their intensity) equally in terms of their potential for $\nu$ production~\citep{Fukuda:2002nf}, and/or examined only the highest-intensity burst(s)~\citep{Thrane:2009fj}. We introduce a novel technique, the Maximum Likelihood Burst Analysis (MLBA), which features some advantages over previous analysis techniques when searching data from neutrino detectors for temporal correlations with astrophysical burst events.  The MLBA provides the versatility to deal with bursts whose intensities vary over many orders of magnitude, and to accommodate correlations between $\gamma$ intensity and $\nu$ emission. This allows us to use the integrated intensity over a large number of bursts instead of only the most intense burst. The results from the MLBA can therefore be employed to test a variety of models in a straightforward manner and, as with previous analysis techniques, can be employed to set limits on the number of neutrino events correlated with the bursts -- and hence indirectly set limits on the fluences.

We begin with a discussion of the SNO detector before discussing two types of astrophysical events, GRB's and solar flares, on which we have chosen to focus.  In section~\ref{sec:MLBA}, we present the MLBA method and motivation.  Next, we examine solar flare and GRB observations in conjunction with the SNO neutrino data set, and give limits for the associated parameter for neutrino fluences from the burst events.  We also show how this can be related to the neutrino fluence limits that have been typically presented in the literature.  In the final section, we search for signals in the SNO data associated with two unusual isolated astrophysical events: the Parkes radio burst~\citep{Lorimer:2007qn} and the SGR 1806-20 magnetar eruption~\citep{Palmer:2005mi}.  In all cases, no significant correlations are seen between the SNO data and the astrophysical sources.  However, limits on neutrino fluences from solar flares and GRB's are improved in the low-energy $\nu$ regime compared to previous analyses in the literature.

\section{The SNO Detector}
\label{sec:detector}
\begin{figure}[tbp]
\includegraphics[width=\columnwidth]{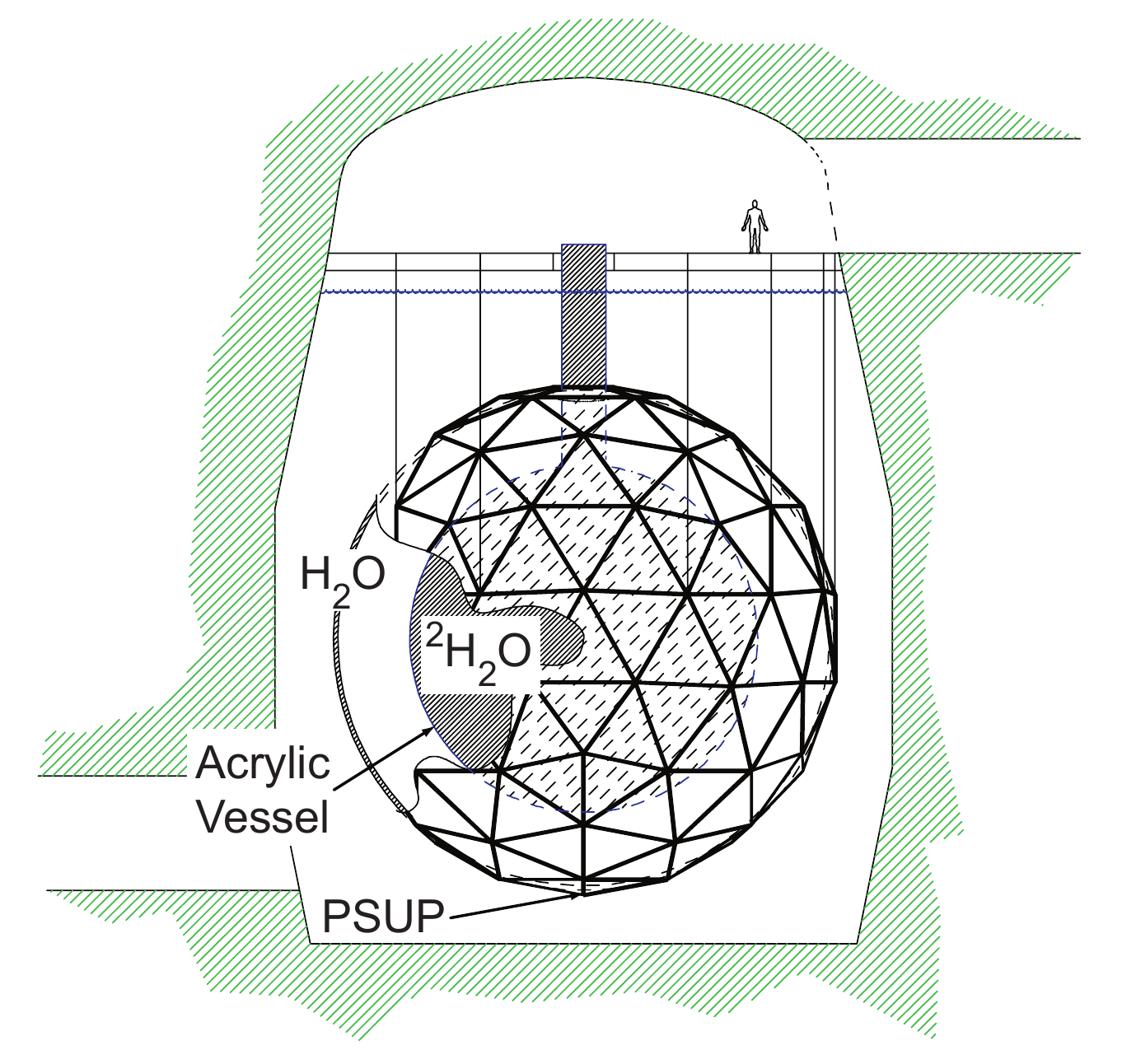}
\caption{(Color online) Schematic diagram of the SNO detector. We used a coordinate system with the center of the detector as the origin, and $z$ direction as vertically upward.}
\label{fig:snodet}
\end{figure}

The SNO detector~\citep{Boger:1999bb}, shown schematically in Figure~\ref{fig:snodet}, consisted of an inner volume containing $10^6$\,kg of 99.92\% isotopically pure heavy water (${\rm ^2H_2O}$, hereafter referred to as \heavywater{}) within a 12 m diameter transparent acrylic vessel (AV). Over $7\times 10^6$\,kg of \water{} between the rock and the AV shielded the \heavywater{} from external radioactive backgrounds. An array of 9456 inward-facing 20 cm Hamamatsu R1408 photomultiplier tubes (PMTs), installed on an 17.8 m diameter stainless steel geodesic structure (PSUP), detected Cherenkov radiation produced in both the \heavywater{} and \water{}. The PMT thresholds were set to 1/4 of the charge from a single photoelectron. The inner $1.7\times 10^6$\,kg of \water{} between the AV and the PSUP shielded the \heavywater{} against radioactive backgrounds from the PSUP and PMTs.

The detector was located in Vale's Creighton mine ($46^\circ 28'30''$\,N latitude, $81^\circ 12'04''$\,W longitude) near Sudbury, Ontario, Canada, with the center of the detector at a depth of 2092\,m (5890$\pm$94 meters water equivalent). At this depth, the rate of cosmic-ray muons entering the detector was approximately three per hour. Ninety-one outward-facing PMTs attached to the PSUP detected cosmic-ray muons. An offline veto based on information from these PMTs significantly reduced cosmogenic backgrounds.

SNO detected low energy neutrinos through the following reactions:
\begin{itemize}
\item $\nu_x + d\rightarrow p + n + \nu_x$, Neutral current (NC),
\item $\nu_e + d\rightarrow p + p + e^-$, Neutrino Charged current (CC),
\item $\bar{\nu}_e + d\rightarrow n + n + e^+$, Anti-Neutrino Charged current (CC-anti),
\item $\nu_x + e^-\rightarrow \nu_x + e^-$, Elastic scattering (ES),
\end{itemize}
where $x$ implies electron, muon or tau neutrinos. The NC reaction is equally sensitive to all three active neutrino and anti-neutrino flavors. This reaction is sensitive down to neutrino energies of approximately 2.2\,MeV, which gives us a lower threshold than many other experiments. The CC reaction is only sensitive to \nue{}s, and the CC-anti reaction is only sensitive to \nuebar{}s. The ES reaction is sensitive to all neutrino and anti-neutrino flavors, but the cross-section for \nue{}s (\nuebar{}s) is approximately six (three) times larger than that for the other flavors. For neutrino energies above approximately 6\,MeV, the CC reaction has a cross-section more than ten times higher than the ES reaction, which provides increased sensitivity for an equivalent target exposure compared to experiments that are only sensitive to the ES reaction.

The recoil electrons from both the ES, CC, and CC-anti reactions were detected directly through their production of Cherenkov light. The total amount of light detected by the PMT array was correlated with the energy of the recoil electron.

The SNO detector operated in three distinct phases distinguished by how the neutrons from the NC interactions were detected. In the \heavywater{} phase, the detected neutrons captured on deuterons in the \heavywater{} releasing a single 6.25 MeV $\gamma$-ray, and it was the Cherenkov light of secondary Compton electrons or $e^+e^-$ pairs that was detected. In the salt phase, $2\times10^3\,{\rm kg}$ of NaCl were added to the \heavywater{}, and the neutrons captured predominantly on $\iso{35}{Cl}$ nuclei, which have a much larger neutron capture cross-section than deuterium nuclei, resulting in a higher neutron detection efficiency. Capture on chlorine also released more energy (8.6 MeV) and yielded multiple $\gamma$-rays, which aided in identifying neutron events. In the NCD phase, an array of proportional counters (the Neutral Current Detection, or NCD, array) was deployed in the \heavywater{}~\citep{Amsbaugh:2007ke}, but some sensitivity remained in the data from the PMT array for neutrons capturing on the \heavywater{} as in the first phase.

Neutrino events observed with the PMT array were selected with the same selection and reconstruction criteria as those in~\citet{Aharmim:2011vm} except the data-set features no high energy cutoff. It consists almost entirely of events below 20 MeV. Table~\ref{tab:time} gives the run times and the rate of events consistent with being neutrinos.
\begin{table*}[htdp]
\begin{center}
\caption{Run times and event rates for the three phases of SNO. The dates are given in modified Julian date (MJD).}
\label{tab:time}
\begin{tabular}{|c|c|c|c|c|c|}
\hline
Phase & On (MJD) & Off (MJD) & Low energy cutoff (MeV)& Livetime (days) & Event rate (day$^{-1}$)\\
\hline
 \heavywater{} & 51484 & 52056 & 5.0& 282.6 & 9.29 \\
Salt  & 52116 & 52879 & 5.5& 385.5 & 8.15 \\
NCD  & 53336 & 54067 & 6.5& 364.4 &  4.09\\
\hline
\end{tabular}
\end{center}
\end{table*}%

\section{Burst Events}

\subsection{Solar Flares}
\label{sec:flare_intro}
The Homestake solar neutrino experiment reported a small excess of events possibly correlated with large solar flares~\citep{Davis:1994jw, Davis:1995wj}, which provided a motivation for subsequent searches. If these events were associated with pion production in the solar atmosphere then these would produce both electron and muon neutrinos with energies up to approximately 50 MeV. The Kamiokande-II collaboration has reported no excess of events associated with large solar flares~\citep{Hirata:1988sx, Hirata_90}. Analysis of the LSD data has also reported no excess of events associated with large solar flares~\citep{Aglietta:1991fv}. Neutrinos could also be produced by low energy beta decay due to excitation in the solar atmosphere. ~\citet{Bahcall:1988vz} has argued that although solar flares could produce neutrinos, they should not be observable in any current detector.

For the coincidence analysis presented here, the solar flare data is taken from HESSI (the High Energy Solar Spectroscopic Imager), which registers radiation ranging from 3~keV to 17~MeV with highly accurate timing.  Extensive data-sets are available from~\citep{HESSI}, with data from over 20000 flares -- the analysis performed in this paper requires simply the time ($t_j$), duration ($\delta t_j$) and intensity ($I_j$) of each flare.  $I_j$ is defined as the total number of registered counts and varies over 10 orders of magnitude. To speed up the analysis we only include events with intensity $I_j > 10^5$ photon counts, which reduces the number of events to 842.
  $\delta t_j$ is typically around 30~minutes, but is also event-dependent. Data from HESSI starts on modified Julian date (MJD) 52330, thus overlapping with the last half of the salt phase and completely with the NCD phase. 

\subsection{$\gamma $-ray Bursters}

\label{sec:grb_intro}
$\gamma $-ray bursters (GRB's) have been known since the 1960's, and an exhaustive review is provided in~\citep{Piran:2004ba}. GRB's produce very large fluxes of $\gamma$ rays over a definite period of time. There are two classes, short and long. Short GRB's emit $90\%$ of the energy within about 100~ms, while long GRB's emit $90\%$ of the energy within about 100~s.  Many models for explaining GRB's have been proposed. Some models suggest a comparatively large flux of high energy (GeV to TeV) $\nu$'s, which might be detectable in (e.g.) ICECUBE. Hence it is reasonable to ask if $\nu$'s could be seen by SNO. The answer is almost certainly not, as fluxes of low-energy (MeV) $\nu$'s should be very low, such that no existing or planned detector could see them from cosmological distances~\citep{Piran:2004ba}.  However, given the uncertainty in the GRB models, it is useful to look for coincidences between GRB's and SNO events. Note that Super-Kamiokande (SK)~\citep{Fukuda:2002nf, Thrane:2009fj} has already searched for such coincidences, but with no significant effect observed.

The ``time of a GRB" is defined as the time at which the burst starts. Since the real nature of GRB's is unknown, there are various possibilities for the timing of the associated $\nu $'s.  In this analysis we have chosen two time windows to cover different possible scenarios:
\begin{enumerate}
\item Assume that the $\nu$'s are produced before the  $\gamma$'s (e.g. a hypernova would have a core collapse followed by a visible outburst after some hours.)  Based on SN 1987a the time difference could be as much as 3 hours. For this, we search for $\nu$'s in an asymmetrical window -3~hrs~$< \delta t < 0$.
\item Assume the initial  $\gamma$'s then trigger a secondary process which emits $\nu$'s. In addition, assuming the $\nu $'s are massive, they would be slightly delayed due to the travel time. For this, we search for $\nu$'s in an asymmetrical window $0 < \delta t < 3$~hrs.  
\end{enumerate}

The GRB data for this analysis is taken from Swift~\citep{Swift}, which is a multi-wavelength observatory dedicated to the study of GRB's, with three instruments that cover $\gamma $-ray, X-ray, UV  and optical frequencies.  The Swift data set starts on MJD 53329, which overlaps completely with the NCD phase, and includes  a total of 190 events.

\section{Maximum Likelihood Burst Analysis}
\label{sec:MLBA}
\subsection{Motivation and Assumptions}
\label{sec:motivation}
Both solar flares and GRB's have hugely varying fluxes.  Clearly, if a burst event is ``strong", it is more likely to have associated $\nu$'s; some bursts would be (hypothetically) too weak to produce a $\nu$ in SNO.  We also need to take into account ordinary solar neutrino (``background") events in SNO.  In addition, the duration of solar flares is highly variable, while various different time windows can be used for GRB's.  A Maximum Likelihood Burst Analysis (MLBA) can adeptly deal with all the above challenges.  
For the MLBA, we suppose the neutrino events in SNO fall into two classes: \begin{enumerate}
\item Random events that arrive at a constant background rate $r_{B}$.
\item Burst events that are associated with some astrophysical trigger (i.e. GRB's or solar flares). These consist of $n_{x}$ events at a time $t_j$ with a 
characteristic  spread $\delta t_j$. In the case of a flare,  $t_j$  would be the start of the flare and $\delta t_j$ its duration.  In the case of a GRB, $t_j$ and $\delta t_j$ would be dictated by the time window being used (see Section~\ref{sec:grb_intro}).
\end{enumerate}
The simplest assumption is that the number of extra neutrino events in SNO (i.e. burst-associated events rather than background) is directly related to the fluence:
\begin{equation}
n_x \left( I \right) = \alpha I.
\label{eqn:alp_def}
\end{equation}  This assumption would be exact if all bursts have identical physical causes but are at varying distances, so that the variations in both $\nu$ and $\gamma$ fluxes are purely geometric. It is also likely to be approximately true in cases where there is a significant variation in intrinsic luminosity, as in the case of Type II supernovae. 

A Maximum Likelihood fit can be used to estimate $\alpha$ by averaging over many bursts.    It is well known that supernovae fall into two very different classes: Type Ia (carbon detonation) supernovae produce very few prompt neutrinos~\citep{Odrzywolek:2010je}, whereas Type II (core collapse) supernovae emit approximately $10^{57}~\nu$'s over a 10~s period.  It is quite possible that GRB's follow a similar pattern.  In such a case it should be noted that we have fitted for the average $\alpha$  over all bursts, and if only some bursts include neutrino emissions then the $\alpha$  for these bursts would be higher.

\subsection{Maximum Likelihood Fitting}
\label{sec:ML_fitting}
We have a total $N_{\rm{SNO}}$ events spread over a total time $T_{tot}$ and a live time $T_{\rm{live}}$. Assuming any actual signal is small, we have a  background rate
\begin{equation}
r_B  = \frac{{N_{{\rm{SNO}}} }}{{T_{{\rm{live}}} }}
\label{E1}
\end{equation}
Note that the background rates $r_B $ for the various phases are given in Table \ref{tab:time}.
Now suppose there is a burst with some characteristic intensity $I_j$  at time $t_j $ which lasts for a time $\delta t_j $. By hypothesis, this will produce $n_x \left( {I_j } \right) = \alpha I_j$ $\nu$'s. In time $\delta t_j $ around the j'th burst, we would expect \begin{equation}
n_j  = w_j \left( {r_B \delta t_j  + \alpha I_j } \right)
\label{E3}
\end{equation}
 events, where  $ w_j$ is a weighting factor for the detector livetime given by
\begin{equation}
w_j  = \frac{1}{\delta t_j}\int_{t_j }^{t_j  + \delta t_j } {\sum\limits_m {H\left( {t - t_m^s } \right)} H\left( {t_m^e  - t} \right)} dt
\label{eqn:weight}
\end{equation}
where $t_m^s \left( {t_m^e } \right)$ are the start (end) times of run $m$, and $H(x)$ is the Heaviside function.  If $w_j < 0.05$ the burst is ignored altogether.
 This implicitly assumes that the events are uniformly distributed over the time window. The actual number of events will form a Poisson distribution, so we would observe $k_j$ events with  a relative probability 
\begin{equation}
P\left( {k_j ,\alpha } \right) = \frac{{n_j^{k_j } }}{{k_j !}}e^{ - n_j } 
\label{E2}
\end{equation}

Hence for all the bursts taken together, the likelihood is:
\begin{equation}
L\left( \alpha  \right) = \prod\limits_{j = 1}^{N_{bursts} } {P\left( {k_j ,\alpha } \right)} 
\label{E4}
\end{equation}

The best-fit value, $\alpha_{fit}$, is found by minimizing $-\ln(L)$ with respect to $\alpha$.   The ${\rm{90\% }}$ upper limits on $\alpha$ are given by
\begin{equation}
-\ln(L(\alpha_{90})) = -\ln(L(\alpha_{fit})) + 0.821762.
\label{eqn:ML_lims}
\end{equation}
It is possible for the fit to give $\alpha _{fit} < 0$, which is obviously unphysical, but in such a case Eq~(\ref{eqn:ML_lims}) can still give us a physical upper limit on $\alpha$. 

An upper limit can also be obtained via Monte Carlo simulations using a Feldman-Cousins style~\citep{Feldman:1997qc} method, which provides a completely independent check on the method.  This involves creating many MC datasets with some fixed $\alpha$ value ($\alpha_{in}$) and running each dataset through the Maximum Likelihood fitter to obtain a result $\alpha_{out}$, repeating this procedure for many different $\alpha_{in}$ values, then plotting the $\alpha_{in}$ vs $\alpha_{out}$ distributions.  We can extract from this an upper limit on $\alpha_{in}$ (i.e. the ``real" value of $\alpha$) for a given value of $\alpha_{out}$ (i.e. the value of $\alpha$ obtained from the Maximum Likelihood fitter). Results for this technique are completely consistent with the MLBA results reported herein.

\subsection{Obtaining Fluence Limits}
\label{sec:method_fluence}
Once a limit on the number of burst-related SNO events is obtained, the next logical step is to turn this into a fluence limit.  Due to the lack of well known spectra, previous results documented in the literature~\citep{Fukuda:2002nf, Hirata:1988sx} express limits on neutrino fluences at the detector in terms of ``Green's function" fluence for mono-energetic neutrinos: 
\begin{equation}
\Phi _\nu  \left( {E_\nu  } \right) = \frac{{N_{90} }}{{\sum\limits_{i = 1}^{n_R } {N_i \int {} \sigma _i \left( {E_\nu  '} \right)\varepsilon _i \left( {E_\nu  '} \right)\delta \left( {E_\nu   - E_\nu  '} \right)dE_\nu  '} }}.
\label{eqn:greens}
\end{equation}
This form takes into account the possibility that there may be several ($n_R $) possible reactions for detecting any particular neutrino flavour. For a given reaction  $\mathit{i}$,  ${\sigma _i \left( E _\nu \right)}$ is the cross-section for neutrinos of energy $E _\nu$, $\varepsilon_i \left( E _\nu \right)$ the detection efficiency, $N_i$ the total number of targets and $N_{90}$ the 90-percent confidence limit on the number of burst-associated neutrinos registered at the detector.  Note that only data from the photomultiplier array in the SNO detector was used, not the data from the independent array of neutron detectors in the NCD phase of the experiment.

To calculate $N_{90}$ in Eq~(\ref{eqn:greens}), we start by taking the fitted value $\alpha _{90}$ (the $90\% $ confidence limit for $\alpha$.) From this, we find the expected number of burst-related $\nu $'s per burst from Eq~(\ref{eqn:alp_def}), weighted by the windowing function in Eq~(\ref{eqn:weight}).  Altogether, this yields: 
\begin{equation}
N_{90}  = \frac{{\alpha _{90} }}{{N_{burst} }}\sum\limits_{j = 1}^{N_{burst} } {I_j w_j } 
\label{N90}
\end{equation}
Note that this does not include systematic uncertainties, which are small compared to the statistical uncertainties.

For each of the reactions we parameterize the low-energy cross-section in Eq~(\ref{eqn:greens}) as 
\begin{equation}
\sigma _i \left( E \right) = \sigma _i^0 \left( {E_\nu   - E_i^0 } \right)^{k_i } 
\end{equation} 
where the values of the threshold energy ${E_i^0}$ are calculated, and $\sigma _i^0$ and $k_i$ are fitted using energies below 20~MeV. Each reaction also has its own $\varepsilon_i (E_\nu  )$, which depends on one or more factors (photomultiplier tube efficiency for electron detection, neutron capture efficiency, etc.) and varies from phase to phase in the SNO experiment.  The number of targets also varies from reaction to reaction. 

\section{Results}

\subsection{Flare Results}
\label{sec:flare_results}
The MLBA results for $\alpha$ for the solar flares are shown in Table~\ref{tab:flare_res}: the larger number of flares during the SNO salt phase is due to the overlap with the solar maximum.  There is no evidence for a signal, for either the SNO salt phase or NCD phase. 
\begin{table*}
\caption{MLBA results for solar flares.  Note that the units of $\alpha$ are [SNO events/HESSI photon counts].  $\alpha_{90}$ refers to the $90\%$ upper confidence bound on $\alpha$, and $N_{90} $ is the $90\% $ upper confidence bound on the average number of flare-associated SNO events per flare.}
\centering
\begin{tabular}{|c|c|c|c|c|c|c|}
\hline
SNO Data & $\alpha _{fit}$ & $\alpha _{90}$ & $\ln[L(\alpha )]-\ln[L(0)]$ & No. Flares&$N_{90} $\\
\hline
Salt Phase	& $2.79 \times 10^{-9}$ &$1.21 \times 10^{-8}$ & 0.13 & 172&0.057\\
NCD Phase & $5.38 \times 10^{-10}$ & $3.44 \times 10^{-9}$ & 0.07 & 94&0.022\\
\hline
\end{tabular}
\label{tab:flare_res}
\end{table*}

\begin{figure}[htbp] 
  \centering
  \includegraphics[width=\columnwidth]{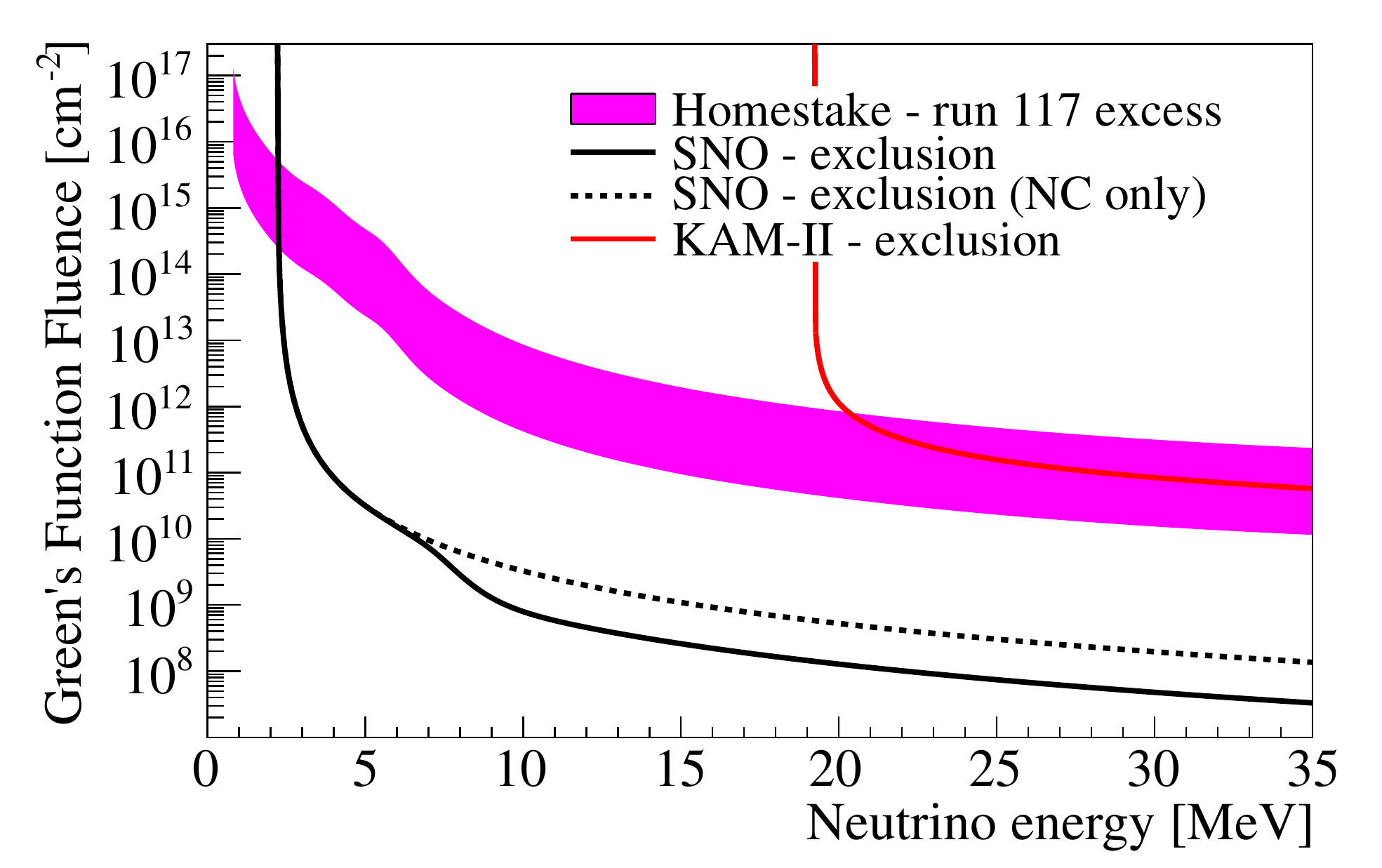} 
\caption{Fluence of neutrinos from solar flares versus neutrino energy. The excess shown for Homestake occurred during run 117, which corresponded to a large solar flare. The Homestake result is obtained assuming the excess in run 117 is attributed to a solar flare.The results for all experiments are calculated assuming pure $\nu_e$ production in the solar atmosphere and include vacuum oscillations during their journey to Earth. The results for Kam II are recalculated from~\citep{Hirata:1988sx} based on this model and assuming a 100\% detection efficiency for scattered electrons above 19\,MeV and 0\% efficiency below this.}
   \label{sens_K}
\end{figure}

Figure \ref{sens_K} shows a comparison of the excess observed in Homestake run 117 in comparison to exclusion limits from the results presented here and KAM II ~\citep{Hirata:1988sx}. Excesses were claimed to be seen in other Homestake runs with similar limits. The results are obtained assuming all the neutrinos are generated as $\nu_e$'s in the Sun's atmosphere, but due to vacuum oscillations the probability of detecting the neutrinos in that state is 0.55. The remaining probability has the neutrinos as either $\nu_\mu$ or $\nu_\tau$. This analysis excludes the Homestake result down to approximately 2.2\,MeV. The KAM II and Homestake results are both obtained from single large bursts, whereas this result is obtained from multiple bursts. Also the Homestake result is obtained with a burst occurring during a run that lasted six days (MJD 48408.89 to 48414.85), whereas this result is obtained with a significantly shorter interval corresponding to the actual flare.

\subsection{GRB Results}
\label{sec:grb_res}
The MLBA results for the GRB's are shown in Table~\ref{tab:GRB_res_swift}.  There is no evidence for a signal in either timing window.  The $90\%$ upper confidence bound on $\alpha$ obtained from the Maximum Likelihood fitting was verified using Monte Carlo simulation and the Feldman-Cousins technique as described at the end of Section~\ref{sec:ML_fitting}.

\begin{table*}
\caption{MLBA results for GRB's, for the Swift dataset (with the SNO NCD data).  Note that units of $\alpha$ are [SNO events $\times$ cm$^2 / 10^{-7}$ erg].  $N_{90} $ is the $90\% $ upper confidence bound on the number of GRB-associated SNO events, based on $\alpha _{90}$.}
\centering
\begin{tabular}{|c|c|c|c|c|c|}
\hline
Timing & $\alpha _{fit}$ & $\alpha _{90} $ &$\ln[L(\alpha _{fit})] - \ln[L(0)]$ & No. GRB's & $N_{90}$\\
\hline
-3 hrs $< \delta t < 0$ & $-2.6 \times 10^{-4}$ & $2.1 \times 10^{-3}$ & 0.013 & 116 & 0.084\\
$ 0 < \delta t <$ 3 hrs & $1.2 \times 10^{-3}$ & $3.9 \times 10^{-3}$ & 0.286 & 116 & 0.164\\
\hline
\end{tabular}
\label{tab:GRB_res_swift}
\end{table*}

\begin{figure}[htbp] 
  \centering
  \includegraphics[width=\columnwidth]{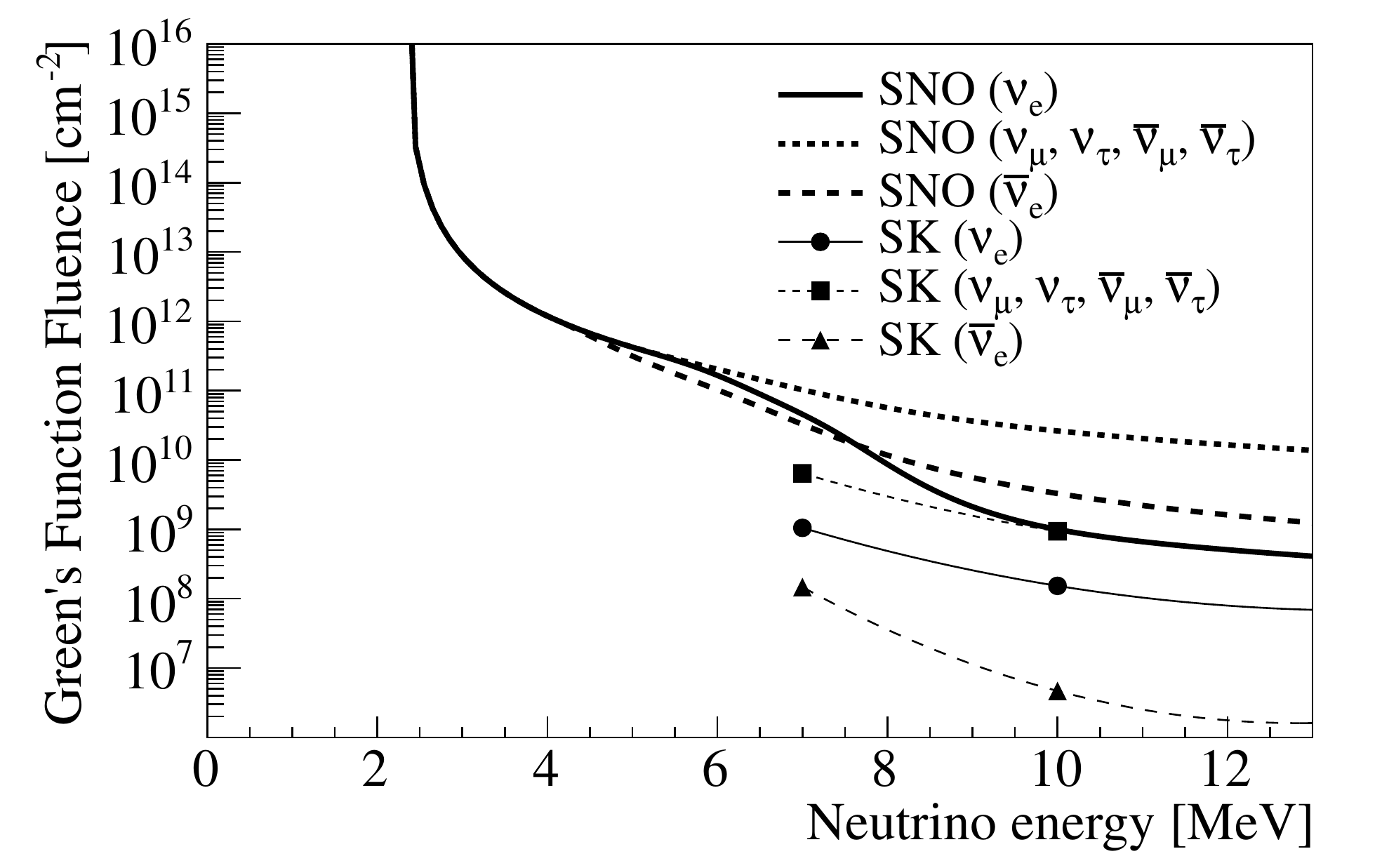} 
\caption{Fluence of neutrinos from GRB versus neutrino energy. The results for SK are obtained from~\citep{Fukuda:2002nf}. The authors do not calculate results below 7\,MeV, but the limits are expected to have a sharp turn up due to the threshold for detecting scattered electrons. The results from SNO presented here are from the NCD phase using SWIFT data with a $-3\,{\rm hrs}<\delta t<0\,{\rm hrs}$ time window. Results from other time windows and phases are similar.}
   \label{fig:grb_fluence}
\end{figure}

To facilitate comparison with the results of SK~\citep{Fukuda:2002nf}, Table~\ref{tab:grb_fluence_greens} converts the Swift results for the NCD phase (-3~hrs$< \delta t <$ 0 timing window) to a ``Green's function" fluence, using  Monte Carlo to obtain the necessary upper limits on the number of GRB-associated events.  These results are plotted in comparison to SK's in Figure~\ref{fig:grb_fluence} for various neutrino flavours.

\begin{table*}
\caption{GRB fluence 90\% CL upper limits: ``Green's function" fluences, using MLBA results for Swift data with the SNO NCD phase (-3 hrs $< \delta t < 0$ hrs timing window.)}
\begin{center}
\begin{tabular}{|c|c|c|c|}
\hline
Energy [MeV] & $\Phi _{\nu _e} \left[ cm^{-2} \right]$
& $\Phi _{\bar \nu _e} \left[ cm^{-2} \right]$ & $\Phi _{\nu _x},\Phi _{\bar \nu _x} \left[ cm^{-2} \right]$ \\
\hline
5 & $3.89\times 10^{11}$ & $2.79 \times 10^{11}$ & 3.92 $\times 10^{11}$ \\
7 & $3.96 \times 10^{10}$ & $3.83 \times 10^{10}$ & $9.64 \times 10^{10}$ \\
9 & $1.92 \times 10^9$ & $4.50 \times 10^9$ & $3.51 \times 10^{10}$ \\
11 & $6.43 \times 10^8$ & $1.44 \times 10^9$ & $1.99 \times 10^{10}$ \\
13 & $4.01 \times 10^8$ & $8.28 \times 10^8$ & $1.35 \times 10^{10}$\\
\hline
\end{tabular}
\end{center}
\label{tab:grb_fluence_greens}
\end{table*}

\section{Other Events}
\label{sec:other}
Two isolated astrophysical events are worth examining.  The first of these events is the ``Parkes Burst"~\citep{Lorimer:2007qn}, a very intense radio signal with a duration of less than 5~ms found in archival data from the Parkes radio telescope. It occurred on 24~August~2001 at UT~19:50:01 (MJD~52145.82640046). The source is unknown, but is believed to be at great distance.  The second is a giant $\gamma $-ray flare from the magnetar SGR~1806-20~\citep{Palmer:2005mi} on 27~December~2004 at UT~21:30:26.65 (MJD~53366.89614178). In this case, there was a precursor burst 143~s before the main burst. 

As well as searching in the SNO neutrino data-set employed in the rest of this paper, we have also searched a different set of SNO events: the ``muon" set employed in~\citep{Aharmim:2009zm}.  The ``muon" set, comprised of 77285 events, is designed to enhance the muon signal rather than the neutrino signal; approximately 500 of the events are likely to be atmospheric neutrinos while the rest are likely muons.  By searching both SNO data-sets, we can look for correlations of either neutrinos or muons with these isolated astrophysical events.  (High energy muon-neutrinos from any astrophysical source would produce muons in the rock, which is confirmed by the observation of upward going muons. It is straightforward to test for SNO muon events that are time-correlated with one of these isolated astrophysical events.)

We have searched for anomalous SNO events in short time-windows  ($\pm 180$\,s) around these two astrophysical events: results are shown in Table~\ref{tab:PM}.  Obviously there is no effect, and the SNO event rate at these times is compatible with the background rate.
\begin{table*}[htdp]
\caption{Isolated astrophysical events.  The number of SNO events observed is in a $\pm 180$\,s window around the event.  The number of SNO events expected is calculated based on the rate of ``ordinary" SNO neutrino or muon events, which here constitute the ``background". Note that due the lower energy and therefore stricter requirements on backgrounds for the neutrino data compared to the muon data, the neutrino data did not include the run corresponding to SGR 1806.}
\begin{center}
\begin{tabular}{|c|c|c|c|c|}
\hline
Astrophysical Event
&SNO Data-set		
&SNO Events Observed		
&SNO Events Expected 		
\\
\hline
Parkes Burst	
&Muon		
&0		
&2.62
\\
Parkes Burst
&Neutrino		
&3
&1.69\\
SGR 1806				
&Muon	
&1
&2.62\\
\hline
\end{tabular}
\end{center}
\label{tab:PM}
\end{table*}

\section{Conclusions}
We can conclude that SNO has found no evidence for low energy neutrinos produced in coincidence with solar flares or GRB's. The MLBA technique allows for a robust search for temporal correlations between events recorded in detectors and astrophysical events of widely varying intensities.  In addition, regarding neutrino fluences from solar flares, SNO has provided limits in the low-energy $\nu$ regime that are improved compared to previous analyses in the literature.

\section{Acknowledgments}
This research was supported by the following. Canada: Natural Sciences and Engineering Research Council, Industry Canada, National Research Council, Northern Ontario Heritage Fund, Atomic Energy of Canada, Ltd., Ontario Power Generation, High Performance Computing Virtual Laboratory, Canada Foundation for Innovation, Canada Research Chairs; US: Department of Energy, National Energy Research Scientific Computing Center, Alfred P. Sloan Foundation; UK: Science and Technology Facilities Council; Portugal: Funda\c{c}\~{a}o para a Ci\^{e}ncia e a Tecnologia. We thank the SNO technical staff for their strong contributions.  We also wish to thank Dr. Richard Hemingway for his encouragement in this work and Dr. Vicki Kaspi for suggesting SGR 1806 as a candidate.

\bibliographystyle{yahapj}
\bibliography{Astropaper36}{}

\begin{thebibliography}{22}
\expandafter\ifx\csname natexlab\endcsname\relax\def\natexlab#1{#1}\fi

\bibitem[{Aglietta {et~al.}(1991)Aglietta, Badino, Bologna, Castagnoli,
  Castellina, {et~al.}}]{Aglietta:1991fv}
Aglietta, M., Badino, G., Bologna, G., {et~al.} 1991,
  \href{http://dx.doi.org/10.1086/170722}{Astrophys.J., 382, 344}

\bibitem[{Aharmim {et~al.}(2005)}]{Aharmim:2005iu}
Aharmim, B., {et~al.} 2005,
  \href{http://dx.doi.org/10.1103/PhysRevD.72.052010}{Phys.Rev., D72, 052010}

\bibitem[{Aharmim {et~al.}(2009)}]{Aharmim:2009zm}
---. 2009, \href{http://dx.doi.org/10.1103/PhysRevD.80.012001}{Phys.Rev., D80,
  012001}

\bibitem[{Aharmim {et~al.}(2010)}]{Collaboration:2009qz}
---. 2010, \href{http://dx.doi.org/10.1088/0004-637X/710/1/540}{Astrophys.J.,
  710, 540}

\bibitem[{Aharmim {et~al.}(2011{\natexlab{a}})}]{Aharmim:2011vm}
---. 2011{\natexlab{a}}, \href{http://arxiv.org/abs/1109.0763}{{\sffamily
  arXiv:1109.0763 [nucl-ex]}}

\bibitem[{Aharmim {et~al.}(2011{\natexlab{b}})}]{Collaboration:2010gx}
---. 2011{\natexlab{b}},
  \href{http://dx.doi.org/10.1088/0004-637X/728/2/83}{Astrophys.J., 728, 83}

\bibitem[{Amsbaugh {et~al.}(2007)Amsbaugh, Anaya, J.b, Bowles, Browne,
  {et~al.}}]{Amsbaugh:2007ke}
Amsbaugh, J.~F., Anaya, J., J.b, B., {et~al.} 2007,
  \href{http://dx.doi.org/10.1016/j.nima.2007.05.321}{Nucl.Instrum.Meth., A579,
  1054}

\bibitem[{Bahcall(1988)}]{Bahcall:1988vz}
Bahcall, J.~N. 1988,
  \href{http://dx.doi.org/10.1103/PhysRevLett.61.2650}{Phys.Rev.Lett., 61,
  2650}

\bibitem[{Boger {et~al.}(2000)}]{Boger:1999bb}
Boger, J., {et~al.} 2000,
  \href{http://dx.doi.org/10.1016/S0168-9002(99)01469-2}{Nucl.Instrum.Meth.,
  A449, 172}

\bibitem[{Davis(1994)}]{Davis:1994jw}
Davis, R. 1994,
  \href{http://dx.doi.org/10.1016/0146-6410(94)90004-3}{Prog.Part.Nucl.Phys.,
  32, 13}

\bibitem[{Davis(1996)}]{Davis:1995wj}
---. 1996,
  \href{http://dx.doi.org/10.1016/0920-5632(96)00263-0}{Nucl.Phys.Proc.Suppl.,
  48, 284}

\bibitem[{Feldman \& Cousins(1998)}]{Feldman:1997qc}
Feldman, G.~J., \& Cousins, R.~D. 1998,
  \href{http://dx.doi.org/10.1103/PhysRevD.57.3873}{Phys.Rev., D57, 3873}

\bibitem[{Fukuda {et~al.}(2002)}]{Fukuda:2002nf}
Fukuda, S., {et~al.} 2002,
  \href{http://dx.doi.org/10.1086/342405}{Astrophys.J., 578, 317}

\bibitem[{Hirata {et~al.}(1988)Hirata, Kajita, Kifune, Kihara, Nakahata,
  {et~al.}}]{Hirata:1988sx}
Hirata, K., Kajita, T., Kifune, T., {et~al.} 1988,
  \href{http://dx.doi.org/10.1103/PhysRevLett.61.2653}{Phys.Rev.Lett., 61,
  2653}

\bibitem[{Hirata {et~al.}(1990)}]{Hirata_90}
Hirata, K.~S., {et~al.} 1990,
  \href{http://dx.doi.org/10.1086/169088}{Astrophys.J., 359, 574}

\bibitem[{Lorimer {et~al.}(2007)Lorimer, Bailes, McLaughlin, Narkevic, \&
  Crawford}]{Lorimer:2007qn}
Lorimer, D.~R., Bailes, M., McLaughlin, M.~A., Narkevic, D.~J., \& Crawford, F.
  2007, \href{http://dx.doi.org/10.1126/science.1147532}{Science, 318, 777}

\bibitem[{Mission(2009)}]{HESSI}
Mission, N. G.~R. 2009, {HESSI}, {Electronic Catalogue}, {URL:
  http://hesperia.gsfc.nasa.gov/hessi/}

\bibitem[{{NASA}(2006)}]{Swift}
{NASA}. 2006, Swift, {Electronic Catalogue}, {URL:
  http://heasarc.gsfc.nasa.gov/docs/swift/swiftsc.html}

\bibitem[{{Odrzywolek, A.} \& {Plewa, T.}(2011)}]{Odrzywolek:2010je}
{Odrzywolek, A.}, \& {Plewa, T.} 2011,
  \href{http://dx.doi.org/10.1051/0004-6361/201015133}{Astron.Astrophys, 529,
  A156}

\bibitem[{Palmer {et~al.}(2005)Palmer, Barthelmy, Gehrels, Kippen, Cayton,
  {et~al.}}]{Palmer:2005mi}
Palmer, D.~M., Barthelmy, S., Gehrels, N., {et~al.} 2005,
  \href{http://dx.doi.org/10.1038/nature03525}{Nature, 434, 1107}

\bibitem[{Piran(2004)}]{Piran:2004ba}
Piran, T. 2004,
  \href{http://dx.doi.org/10.1103/RevModPhys.76.1143}{Rev.Mod.Phys., 76, 1143}

\bibitem[{Thrane {et~al.}(2009)}]{Thrane:2009fj}
Thrane, E., {et~al.} 2009,
  \href{http://dx.doi.org/10.1088/0004-637X/697/1/730}{Astrophys.J., 697, 730}

\end{thebibliography}

\end{document}